\documentclass[preprint,aps]{revtex4}
\usepackage{amsmath}    
\usepackage{amsfonts}  
\usepackage{amssymb}
\usepackage{graphicx}   
\usepackage{graphicx}
\usepackage{dcolumn}
\usepackage{listings}
\usepackage{xcolor}
\usepackage{hyperref}
\usepackage{fancyhdr}

\begin{document}

\title{Constructing a chaotic system with any number of equilibria}

\author{Xiong Wang\footnote{Corresponding author.}}
\email{wangxiong8686@gmail.com}
\author{Guanrong Chen}
\email{eegchen@cityu.edu.hk}
\affiliation{Department of Electronic Engineering, City University of Hong Kong, Hong Kong SAR, China\\ 
}

\begin{abstract}

In the chaotic Lorenz system, Chen system and R\"ossler system, their equilibria are unstable and the number of the equilibria are no more than three. This paper shows how to construct some simple chaotic systems that can have any preassigned number of equilibria. First, a chaotic system with no equilibrium is presented and discussed. Then, a methodology is presented by adding symmetry to a new chaotic system with only one stable equilibrium, to show that chaotic systems with any preassigned number of equilibria can be generated. By adjusting the only parameter in these systems, one can further control the stability of their equilibria. This result reveals an intrinsic relationship of the global dynamical behaviors with the number and stability of the equilibria of a chaotic system.\\

\end{abstract}

\maketitle

\noindent {\bf Thus far, one typically calculates the number and determines the corresponding stability of the equilibria of a given chaotic system, but seldom considers how to generate a desirable chaotic system with a preassigned number and preferred stability of equilibria. This article shows that, for 3D autonomous systems, the latter can be done effectively in a fairly simple way. This helps better understand the intrinsic relationship between the equilibria and dynamics of a chaotic system.}

\section{Introduction}

In chaos theory, it is important to study the stability of the equilibria of an autonomous dynamical system. For a dynamical system described by a set of autonomous ordinary differential equations (ODEs), $\dot{\textbf{x}}=f(\textbf{x}),\,\textbf{x}\in R^n\ $, if $f(\textbf{x}_{\rm e})=0$ has real solution then $\textbf{x}_{\rm e}$ is called the \emph{equilibrium\,} of this dynamical system. An equilibrium is said to be \emph{hyperbolic\,} if all eigenvalues of the system Jacobian matrix have non-zero real parts. A hyperbolic equilibrium for three-dimensional (3D) autonomous system can be a \emph{node}, \emph{saddle}, \emph{node-focus\,} or \emph{saddle-focus}. For 3D autonomous hyperbolic type of dynamical systems, a commonly accepted criterion for proving the existence of chaos is due to \v{S}i'lnikov
\cite{Silinkov1,Silinkov2,Silinkov3,Silinkov5,Silinkov6chenEXTENDED}.

It has also been noticed that although most chaotic systems are of hyperbolic type, there are still many others that are not so. For
non-hyperbolic type of chaotic systems, they usually do not have saddle-focus equilibria, such as those found by Sprott
\cite{sprott1993,sprott1994,sprott1997,sprottlinz2000}.

The well-known Lorenz system \cite{lorenz63} and also Chen system
\cite{chen1999,ueta2000} both have two unstable saddle-foci and one unstable node. They can generate a two-wing butterfly-shaped chaotic attractor. Near the center of the two wings, there lies one an unstable saddle-focus. In recent years, some chaotic systems are found which can generate three-wing, four-wing, and even multi-wing attractors. Observe that, typically, for those symmetrical four-wing attractors, near the center of their wings there also lies one unstable saddle-focus. This common feature may imply that the number of equilibria basically determines the shape of a multi-wing attractor. Therefore, it is interesting to ask: Is it possible to generate a chaotic system with an arbitrarily preassigned number of equilibria? Is the number of equilibria always determinate the shape of an attractor?

Furthermore, regarding the stability of the equilibria, recall that recently Yang and Chen found a chaotic system with one saddle and two stable node-foci \cite{yang2008}, and an unusual 3D autonomous quadratic Lorenz-like chaotic system with only two stable node-foci \cite{yang2010}. Moreover, Wang and Chen found an interesting chaotic system with only one stable node-focus \cite{onestable2011}. Thus, another interesting question is whether it is possible for a chaotic system to have two, or three, or even an arbitrarily large number of stable/unstable equilibria?

This paper attempts to answer these questions by showing a novel example of a chaotic system with no equilibrium, and several other examples of chaotic systems which have any preassigned number
of equilibria. Meanwhile it is shown that, by adjusting the only parameter in these systems, one can also control the stability of these equilibria.

\section{Chaotic system with no equilibrium}

First, a 3D chaotic autonomous system system is introduced:
\begin{equation}\label{noeq}
\left\{
\begin{array}{l}
\dot{x}=y\\
\dot{y}=z \\
\dot{z}=-y+3y^{2}-x^{2}-xz-a\,,
\end{array}
\right.
\end{equation}

\begin{figure}
\includegraphics[width=16cm]{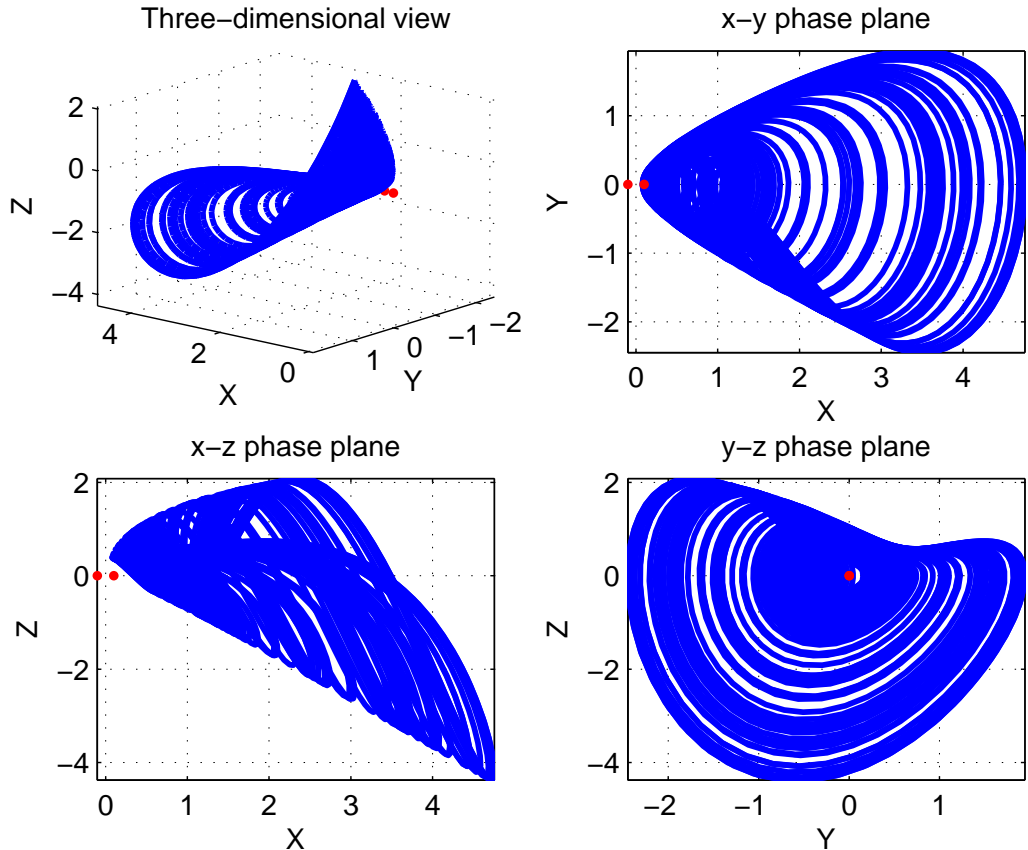}
\caption{(Color online) The chaotic attractor of system (\ref{noeq}), which has two symmetrical equilibria when $a=0.01$: 3D views on the $x$-$y$ plane, $x$-$z$ plane and $y$-$z$ plane.} \label{01}
\end{figure}

\begin{figure}
\includegraphics[width=16cm]{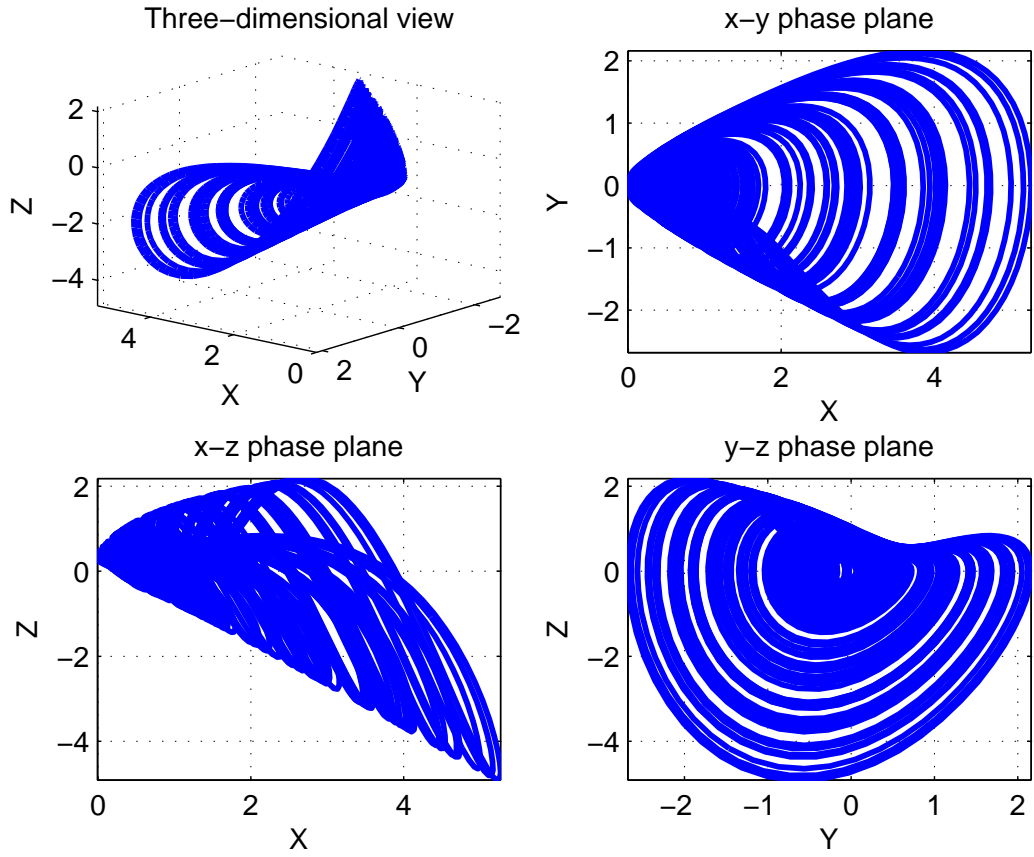}
\caption{(Color online) The chaotic attractor of system \ref{noeq}, which has no equilibrium when $a=-0.05$; 3D views on the $x$-$y$ plane, $x$-$z$ plane and $y$-$z$ plane.} \label{02}
\end{figure}

\begin{figure}
\centering
\includegraphics[width=10cm]{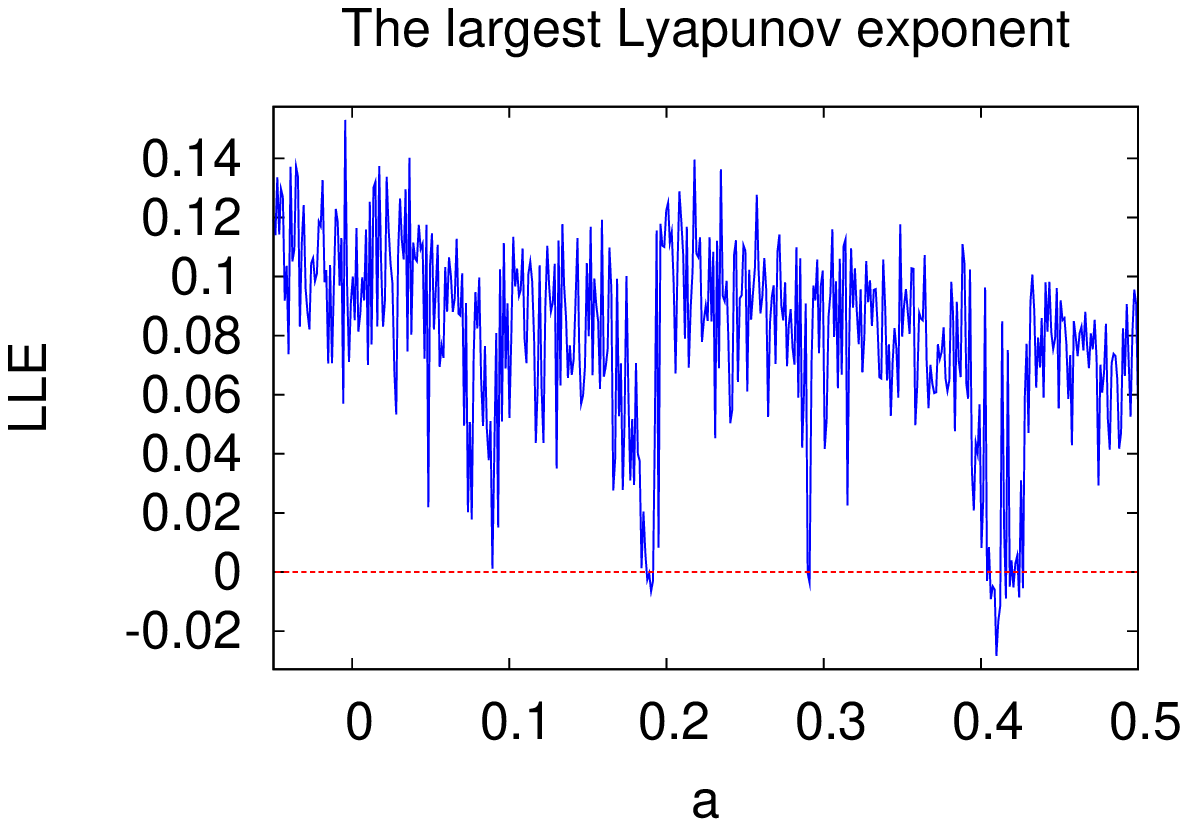}
\caption{(Color online) The largest Lyapunov exponent of system
(\ref{noeq}) with respect to the parameter $a$. }\label{03}
\end{figure}

When $a>0$, this system has two symmetrical equilibria:
$(\sqrt{a},0,0)$ and $(-\sqrt{a},0,0)$, as shown by Fig. \ref{01}. When $a=0$, these two symmetrical equilibria merge into one, the origin $(0,0,0)$. When $a<0$, there is no equilibrium in this system, but still, the system can generate a chaotic attractor, as shown in Fig. \ref{02}. The largest Lyapunov exponent with respect to the parameter $a$ is shown in Fig. \ref{03}, which convincingly implies that the system is chaotic.

\section{A modified Sprott E system with one stable equilibrium}

Now, a chaotic system with only one equilibrium, a stable node-focus, is introduced, which was reported in \cite{onestable2011}. 

This system was found by modifying the Sprott E system, as follows:
\begin{equation}\label{wangeq}
\left\{
\begin{array}{l}
\dot{x}=yz+a\\
\dot{y}=x^{2}-y \\
\dot{z}=1-4x\,.
\end{array}
\right.
\end{equation}

When $a=0$, it is the Sprott E system \cite{sprott1997}; when
$a\neq0$, however, the stability of the single equilibrium is
fundamentally different.

Specifically, when $a>0$, system (\ref{wangeq}) possesses only one stable equilibrium:
\begin{equation}
P \left( x_{E},y_{E},z_{E} \right)
=\left(\frac{1}{4},\frac{1}{16},-16\,a\right)\,.
\end{equation}
\begin{figure}
\includegraphics[width=16cm]{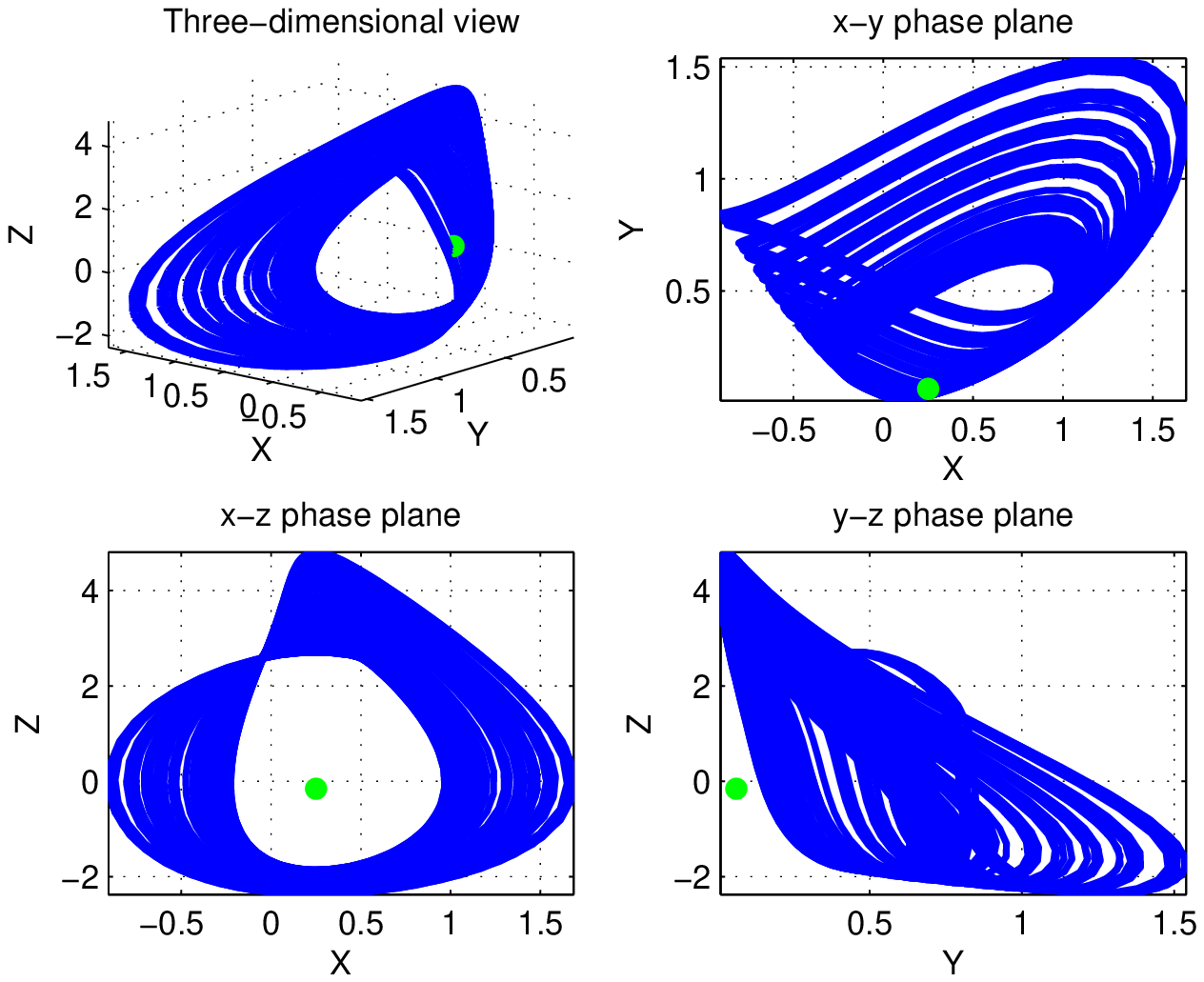}
\caption{(Color online) The chaotic attractor of system (\ref{wangeq}), when $a=0.006$; 3D views on the $x$-$y$ plane, $x$-$z$ plane and $y$-$z$ plane.} \label{0063D}
\end{figure}

Some numerical calculation results are shown in Table 1.
\begin{table}[h]
\caption{Jacobian eigenvalues of system (\ref{wangeq})}
{\begin{center}\begin{tabular}
{|c|c|} \hline
$a$&  Jacobian eigenvalues\\
\hline
\, $0$&$-1$,$ \pm0.5i$\, \\
\, $0.006$&$-0.9607$, $-0.0197\pm0.5098i$\, \\
\, $0.022$&$-0.8458, -0.0771\pm0.5382i$\, \\
\hline
\end{tabular}\end{center}}
\end{table}

Interestingly, this system (\ref{wangeq}) can generate a one-scroll chaotic attractor, as shown in Fig. \ref{0063D}.

In the following, this system (\ref{wangeq}) is further modified by imposing some kind of symmetry onto it, to have different numbers of equilibria while keeping this system chaotic.

\section{Chaotic system with two equilibria}

Rewrite system (\ref{wangeq}) in terms of $u,v,w$ as follows:
\begin{equation}\label{wangeq2uvw}
\left\{
\begin{array}{l}
\dot{u}=vw+a\\
\dot{v}=u^{2}-v \\
\dot{w}=1-4u\,.
\end{array}
\right.
\end{equation}

From Fig. \ref{0063D}, one can see that the $y$-axis does not
intersect with the attractor in system (\ref{wangeq}). So, one may try to add a $y$-axis rotation symmetry to this system, as detailed below.

Consider the following simple coordinate transformation:
\begin{equation}\label{tran2}
\left\{
\begin{array}{l}
u = x^2-z^2\\
v = y\\
w = 2xz\,.
\end{array}
\right.
\end{equation}%
This transformation can add a $y$-axis rotation symmetry, $\mathbb{R}_{y}(\pi)$, to the original system, because for each $(u,v,w)$ there are two points $(\pm x,\pm y,\pm z)$ corresponding to $(u,v,w)$.

After the above transformation, the system becomes
\begin{equation}\label{chaos2}
\left\{
\begin{array}{l}
\dot{x}=\frac{1}{2}{\frac{z+2y{x}^{2}z+xa-4{x}^{2}z+4{z}^{3}}{{x}^{2}+{z}^{2}}}\\
\dot{y}=\left( {x}^{2}-{z}^{2} \right) ^{2}-y\\
\dot{z}=-\frac{1}{2}{\frac{2yx{z}^{2}+za-4x{z}^{2}-x+4{x}^{3}}{{x}^{2}+{z}^{2}}}\,.
\end{array}
\right.
\end{equation}

The new system (\ref{chaos2}) possesses two symmetrical equilibria, which are independent of the parameter $a$:
$P1(\frac{1}{2},\frac{1}{16},0)$ and
$P1(-\frac{1}{2},\frac{1}{16},0)$.

System (\ref{chaos2}) is not globally but only locally topologically equivalent to the original system (\ref{wangeq}), however. Yet one can control the stability of these equilibria by
adjusting the parameter $a$, so that the stability remains the same as the original system (\ref{wangeq}) which, when $a<0$ are unstable and when $a>0$ are stable.

By linearizing system (\ref{wangeq}) at
$P1(\frac{1}{2},\frac{1}{16},0)$, one obtains the Jacobian
\begin{eqnarray}
J\,\Big|_O =\left[
  \begin{array}{ccc}
  -2a & 0 & \frac{1}{16} \\
  \frac{1}{2} & -1 & 0 \\
  -4 & 0 & -2a \\
  \end{array}
  \right]\,,
\end{eqnarray}
whose characteristic equation is
\[
 {\rm det}(\lambda I-J|_O)
 =\lambda^3+(1+4a)\lambda^2+\Big(4a^2+4a+\frac{1}{4}\Big)\lambda
 +4a^2+\frac{1}{4}=0\,,
\]
which yields
\begin{eqnarray}
\lambda_1&=&-1<0,\nonumber\\
\lambda_2&=&-2a+0.5i,\nonumber\\
\lambda_3&=&-2a-0.5i\,.\nonumber
\end{eqnarray}

\begin{figure}
\centering
\includegraphics[width=16cm]{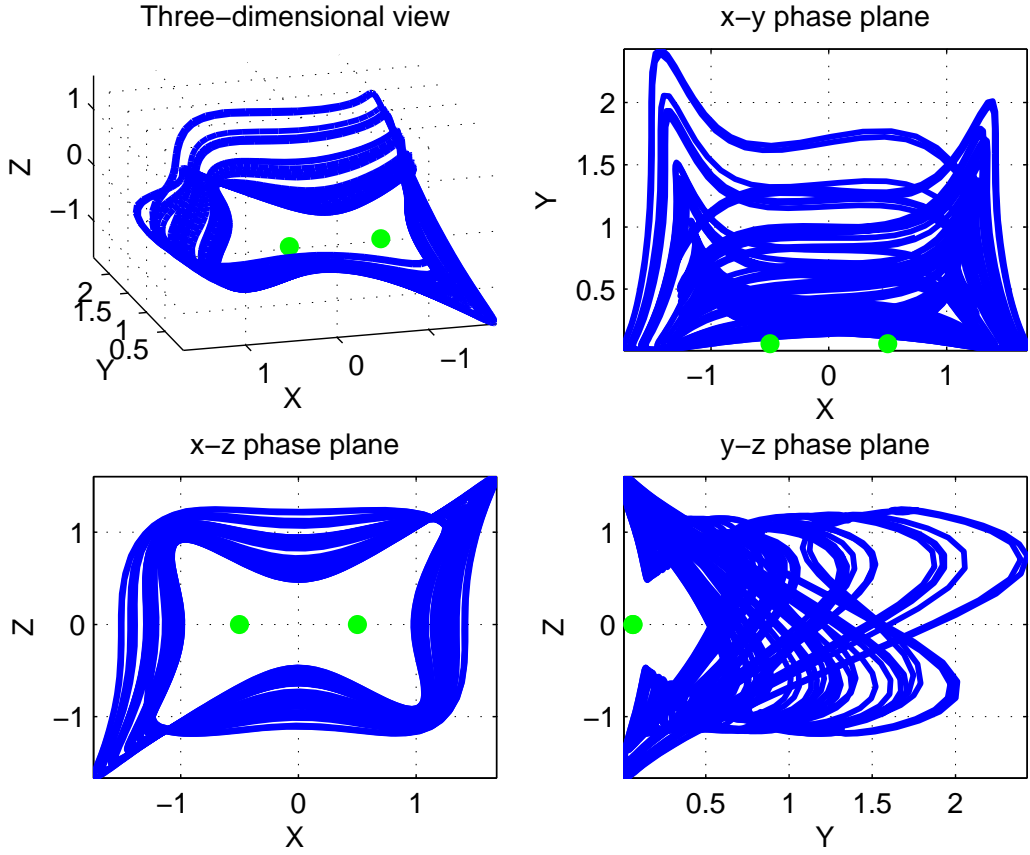}
\caption{(Color online) The new two-petal chaotic attractor with
stable equilibria when $a=0.003$.}\label{2s}
\end{figure}

\begin{figure}
\centering
\includegraphics[width=16cm]{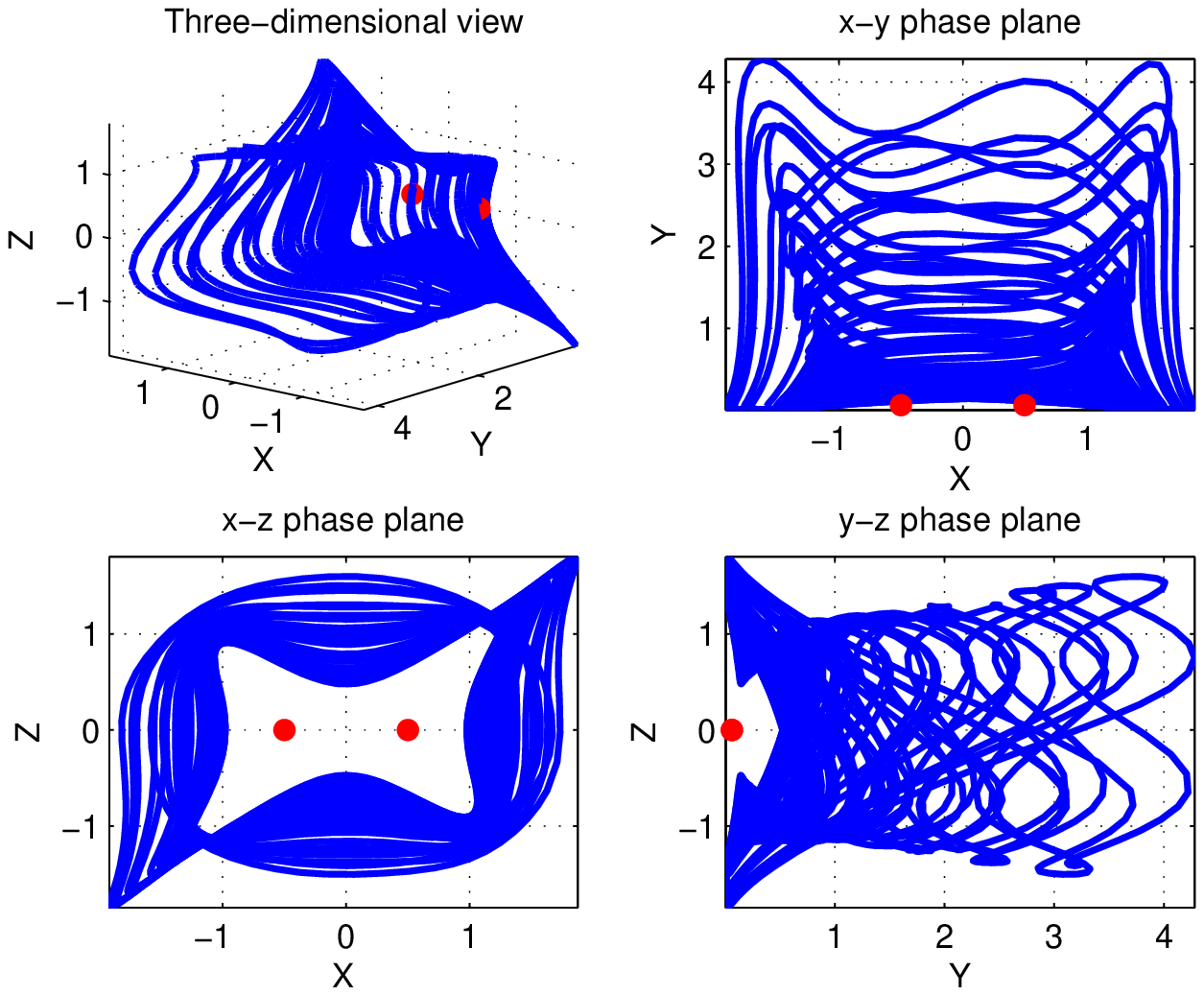}
\caption{(Color online) The new two-petal chaotic attractor with
unstable equilibria when $a=-0.01$.}\label{2us}
\end{figure}

System (\ref{chaos2}) can generate a symmetrical two-petal chaotic attractor, as shown in Fig. \ref{2s} and Fig. \ref{2us}, respectively.

\begin{figure}
\includegraphics[width=10cm]{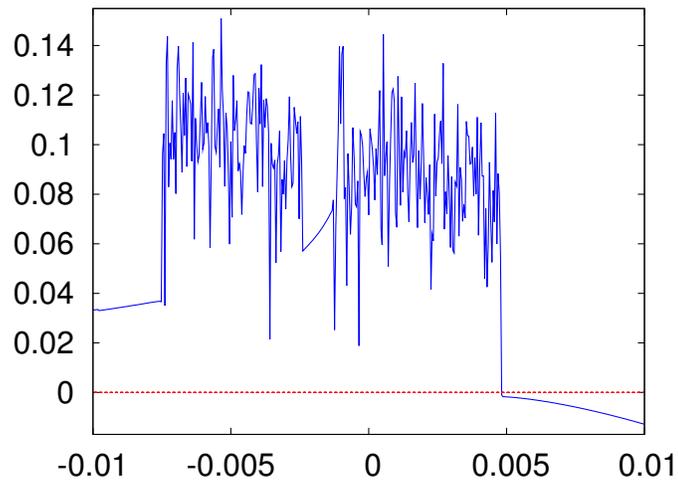}
\caption{(Color online) The largest Lyapunov exponent of system
(\ref{chaos2}) with respect to parameter $a$. }\label{2lle}
\end{figure}

Numerical calculation of the largest Lyapunov exponent of the the system indicates the existence of chaos for some particular values of parameter $a$, as shown in Fig. \ref{2lle}.

\section{Chaotic system with three equilibria}

Similarly, consider the following transformation:
\begin{equation}\label{tran3}
\left\{
\begin{array}{l}
u = x^3-3xz^2\\
v = y\\
w = 3x^2z-z^3\,.
\end{array}
\right.
\end{equation} %
It can add a $y$-axis rotation symmetry, $\mathbb{R}_{y}(\frac{2}{3}\pi)$, to the original system, because for each $(u,v,w)$ there are three symmetrical points corresponding to it.

After the above transformation, the system becomes
\begin{equation}\hspace{-20pt}\label{chaos3}
\left\{\hspace{-5pt}
\begin{array}{l}
\dot{x}=\frac{1}{3}{\frac
{3\,{x}^{4}zy-4\,{x}^{2}{z}^{3}y+{x}^{2}a-8\,{x}^{4}z+2\,z
x+24\,{x}^{2}{z}^{3}+{z}^{5}y-{z}^{2}a}{2\,{x}^{2}{z}^{2}+{x}^{4}+{z}^
{4}}} \\
\dot{y}=(x^3-3xz^2)^2-y \\
\dot{z}=-\frac{1}{3}{\frac
{6\,{z}^{2}{x}^{3}y-2\,{z}^{4}xy+2\,zxa+4\,{x}^{5}-{x}^{2}-16\,
{z}^{2}{x}^{3}+{z}^{2}+12\,{z}^{4}x}{2\,{x}^{2}{z}^{2}+{x}^{4}+{z}^{4}}}\,.
\end{array}
\right.
\end{equation}\label{eq3}

This system possesses three symmetrical equilibria, which are
dependent on the parameter $a$. The analytical expression are too
long to write out here, so only some numerical results are shown in Table \ref{3e}.

\begin{table}[htbp]\caption{Equilibria and Jacobian eigenvalues of system (\ref{chaos3})}
\begin{tabular}{|c|c|p{150pt}|c|}
\hline & $a$ & Equilibria &
Jacobian eigenvalues \\
\hline Unstable case & 
\, $-0.01$\, & P1=(0.6550,0.0625,0.1258) \par
P2=(-0.2186,0.0625,-0.6300) \par P3=(-0.4365,0.0625,0.5044)&
$-1.0617,0.0308\pm 0.4843i$\, \\
\hline Stable case & 
\, 0.01 & P1=(0.6550,0.0625,-0.1258) \par
P2=(-0.2186,0.0625,0.6300) \par P3=(-0.4365,0.0625,-0.5044) & $-0.9334,-0.0333\pm 0.5165i$ \, \\
\hline
\end{tabular}
\label{3e}
\end{table}

\begin{figure}
\centering
\includegraphics[width=16cm]{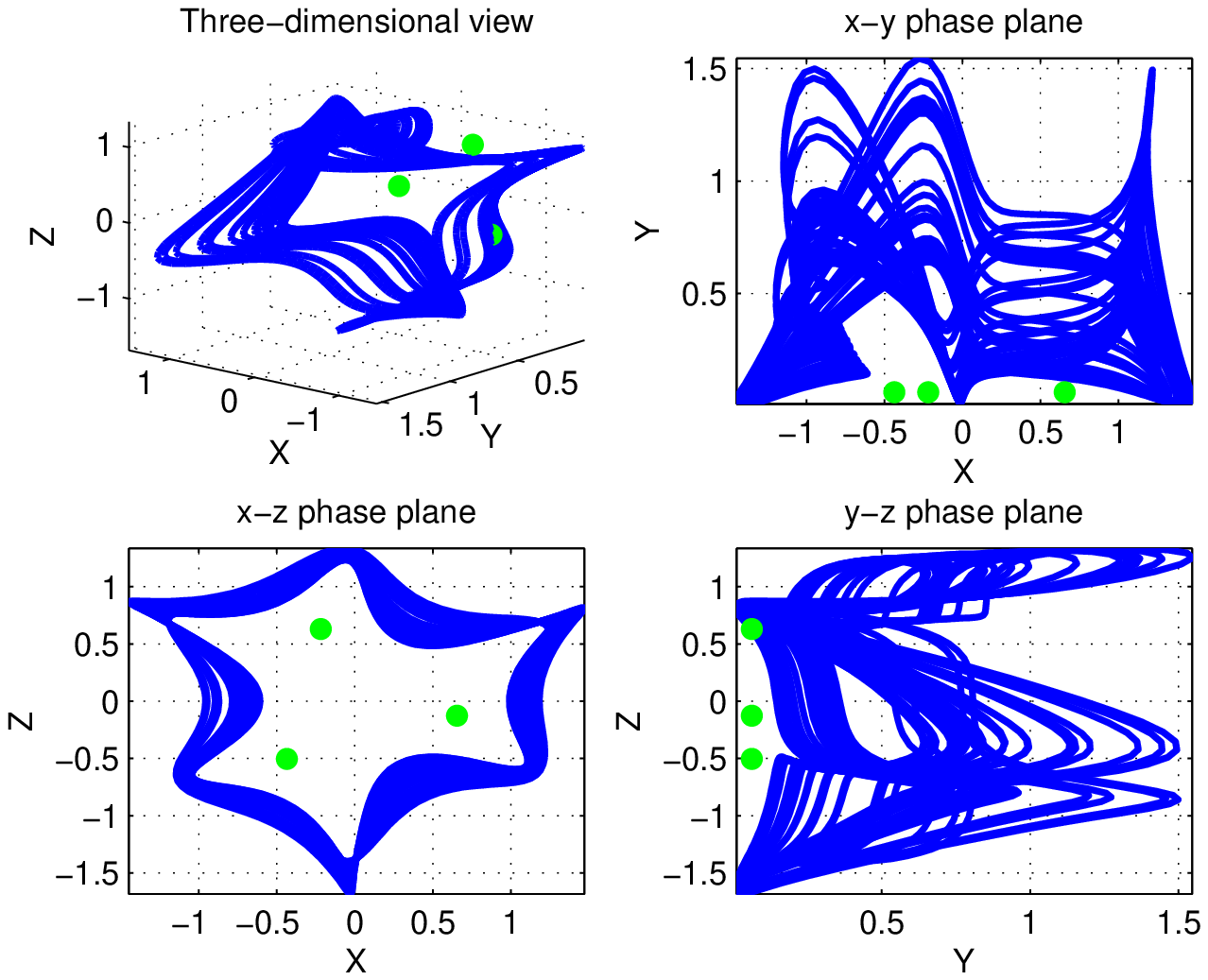}
\caption{(Color online) Chaotic attractor of system (\ref{chaos3}) with three stable symmetrical equilibria when $a=0.01$.}\label{3s}
\end{figure}

\begin{figure}
\centering
\includegraphics[width=16cm]{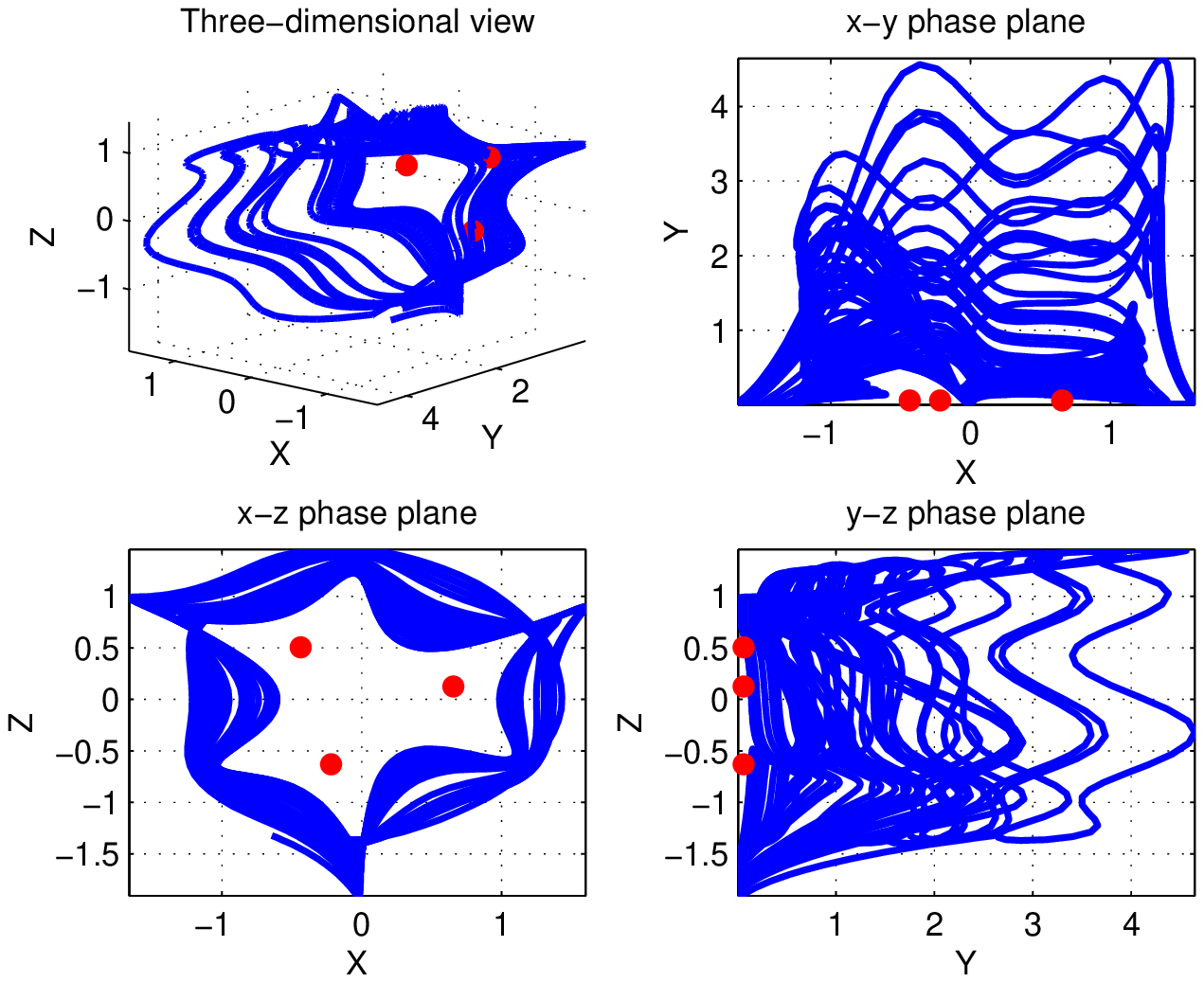}
\caption{(Color online) Chaotic attractor of system (\ref{chaos3}) with three unstable symmetrical equilibria when $a=-0.01$.}\label{3us}
\end{figure}

The system has a symmetrical three-petal chaotic attractor, as shown in Fig. \ref{3s} and Fig. \ref{3us}, respectively. 

\begin{figure}
\includegraphics[width=10cm]{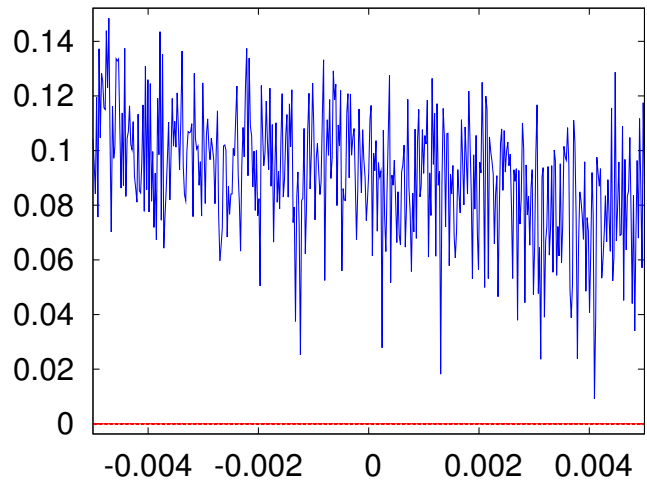}
\caption{(Color online) The largest Lyapunov exponent of system (\ref{chaos3}) with respect to the parameter $a$. }\label{3lle}
\end{figure}

Numerical calculation of the largest Lyapunov exponent indicates the existence of chaos for some particular values of parameter $a$, as shown in Fig. \ref{3lle}.

\section{Chaotic system with any number of equilibria}

Theoretically, one can use the transform $(x+iz)^{n}=(u+iv)$ to
obtain a new chaotic system with $n$ equilibria. The following is such an example with five equilibria.

\begin{figure}
\centering
\includegraphics[width=16cm]{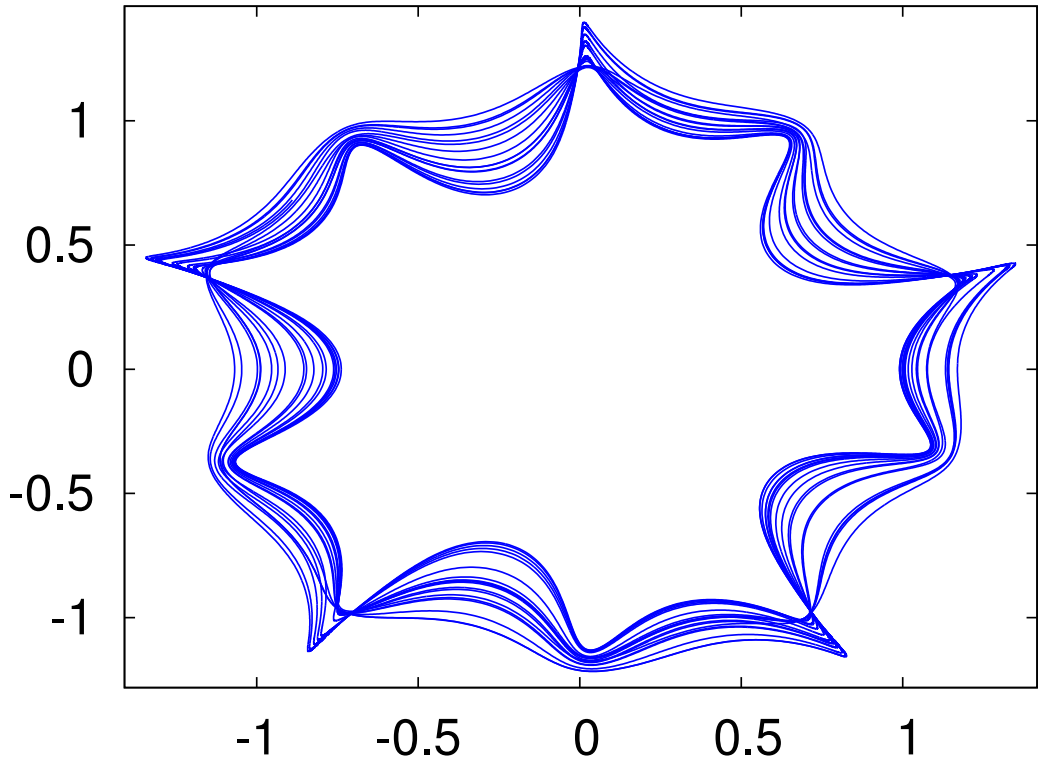}
\caption{(Color online) Chaotic attractor of the new system with five symmetrical equilibria when $a=0$.}\label{5}
\end{figure}

Consider the following transformation:
\begin{equation}\label{tran5}
\left\{
\begin{array}{l}
u = x^5-10x^3z^2+5xz^4\\
v = y\\
w = 5x^4z-10x^2z^3+z^5\,.
\end{array}
\right.
\end{equation} %
It can add a $y$-axis rotation symmetry, $\mathbb{R}_{y}(\frac{2}{5}\pi)$, to the original system. The new equations are too long to write out, so are omitted here. The symmetrical attractor generated with parameter $a=0$ is shown in in Fig. \ref{5}.

\section{Discussions}

The Hartman-Grobman Theorem is an important theorem in ODE systems theory. It is about the \emph{local\,} behavior of an autonomous dynamical system in the neighborhood of a hyperbolic equilibrium, stating that the behavior of the dynamical system near the hyperbolic equilibrium is qualitatively the same as (i.e., topologically equivalent to) the behavior of its linearization near this equilibrium point.

Notice, however, that the new systems discussed in this paper have chaos, which is a \emph{global\,} behavior, although such a system has only one hyperbolic equilibrium point. In other words, all system flows locally converge to the stable equilibrium, but they are chaotic globally. This interesting phenomenon is worth further studying, both theoretically and experimentally, to further reveal the intrinsic relationship between the local stability of an equilibrium and the global complex dynamical behaviors of a chaotic system.

\section*{Acknowledgement}

This research was supported by the Hong Kong Research Grants Council under the GRF Grant CityU1114/11E.

\end{document}